\documentclass[twocolumn,apex]{jjap3_arXiv}
\usepackage{txfonts}
\usepackage{bm}
\usepackage{braket}

\title{First-principles study of electric-field-induced topological phase transition in one-bilayer Bi(111)}%
\author{Hikaru Sawahata$^{1}$\thanks{E-mail: sawahata@cphys.s.kanazawa-u.ac.jp}, Naoya Yamaguchi$^1$, Hiroki Kotaka$^2$, Fumiyuki Ishii$^3$\thanks{E-mail: ishii@cphys.s.kanazawa-u.ac.jp}}
\inst{
$^1$
Graduate School of Natural Science and Technology, Kanazawa University, Kanazawa 920-1192, Japan
\\
$^2$
Elements Strategy Initiative for Catalysts and Batteries (ESICB), Kyoto University, Kyoto 615-8245, Japan
\\
$^3$
Faculty of Mathematics and Physics, Kanazawa University, Kanazawa 920-1192, Japan
}
\abst{
Using first-principles calculations, we found the topological phase transition induced by electric fields in one-bilayer Bi(111). The bandgap decreased with increasing electric field strength, and it is closed at 2.1 V/{\AA}. For fields exceeding 2.1 V/{\AA}, the bandgap increased with increasing electric field strength, reaching 0.34 eV at 4.0 V/{\AA}. We computed the $Z_{2}$ invariant that characterizes topological insulator phases. As results, one-bilayer Bi(111) showed a topological phase transition induced by the electric field, from the topological insulator phase to the trivial insulator phase through a Dirac semimetal. This topological phase transition could be applied to novel devices.
}

\begin{document}
\maketitle

Topological insulators have attracted much research attention for their possible applicability to novel devices\cite{Hasan_Colloquium_2010}. Topological insulator is a non-trivial insulator phase caused by time-reversal symmetry and spin-orbit interactions. The phase has stable metallic edge states protected by time-reversal symmetry; therefore, the metallic edge states are robust against non-magnetic impurities. In addition, the edge states have dissipation-free spin currents. Because of the lack of back-scattering in the edge spin current, topological insulator could be applied in the materials of novel devices. \par

In order to achieve on/off electrical switching of the edge spin current, i.e., to create a trivial-to-topological switching device, 
the topological phase transition between the topological insulator phase and trivial-insulator phase must be controlled, which requires closing the bandgap.
Topological phase transitions induced by electric fields have been predicted for the two-dimensional material of phosphorene. 
It could be applied as a novel transistor using 
the topological phase transitions from trivial insulator to topological insulator, 
where the 0.1 eV bandgap of four-layer phosphorene closes at electric field $E = 0.3$ V/{\AA} reportedly\cite{Liu_Switching_2015}.

\par

It is worth investigating the materials in topological-to-trivial electrical switching devices. 
One-bilayer Bi(111) is one such candidate material. It could be formed on Si(111)  experimentally\cite{Nagao_Nanofilm_2004}. It is reported theoretically\cite{Murakami_Quantum_2006,Wada_Localized_2011}  and experimentally\cite{Hirahara_Interfacing_2011,Yao_Topologically_2016}  as a topological insulator.
The one-bilayer Bi(111) under electric fields is predicted as topological insulator states at $E < 0.8$ V/{\AA}\cite{Chen_Robustness_2013}
and the bandgap of Bi(111) decreases from 0.3 eV to 0.1 eV at $E = 1.5$ V/{\AA}\cite{Kotaka_rashba}.
\par
In this study, we investigate the electric field effect on one-bilayer Bi(111) film,
where electric field is applied up to 4.0 V/{\AA}.
We calculate the electric-field dependence of the band structure and the topological invariant $Z_{2}$ on one-bilayer Bi(111) film. To compute the $Z_{2}$ invariant, we use the Wannier function center (WFC)\cite{Soluyanov_Computing_2011,Yu_Equivalent_2011}
 and the lattice Chern number (LCN)\cite{Fukui_Quantum_2007,Feng_First_2012} because the Parity method\cite{Fu_Topological_2007}  cannot be applied to systems with electric fields. In this study, one-bilayer Bi(111) shows a topological phase transition from the topological insulator phase to a trivial insulator phase under the application of electric fields.

\begin{figure}[htbp]
\includegraphics[width=\columnwidth]{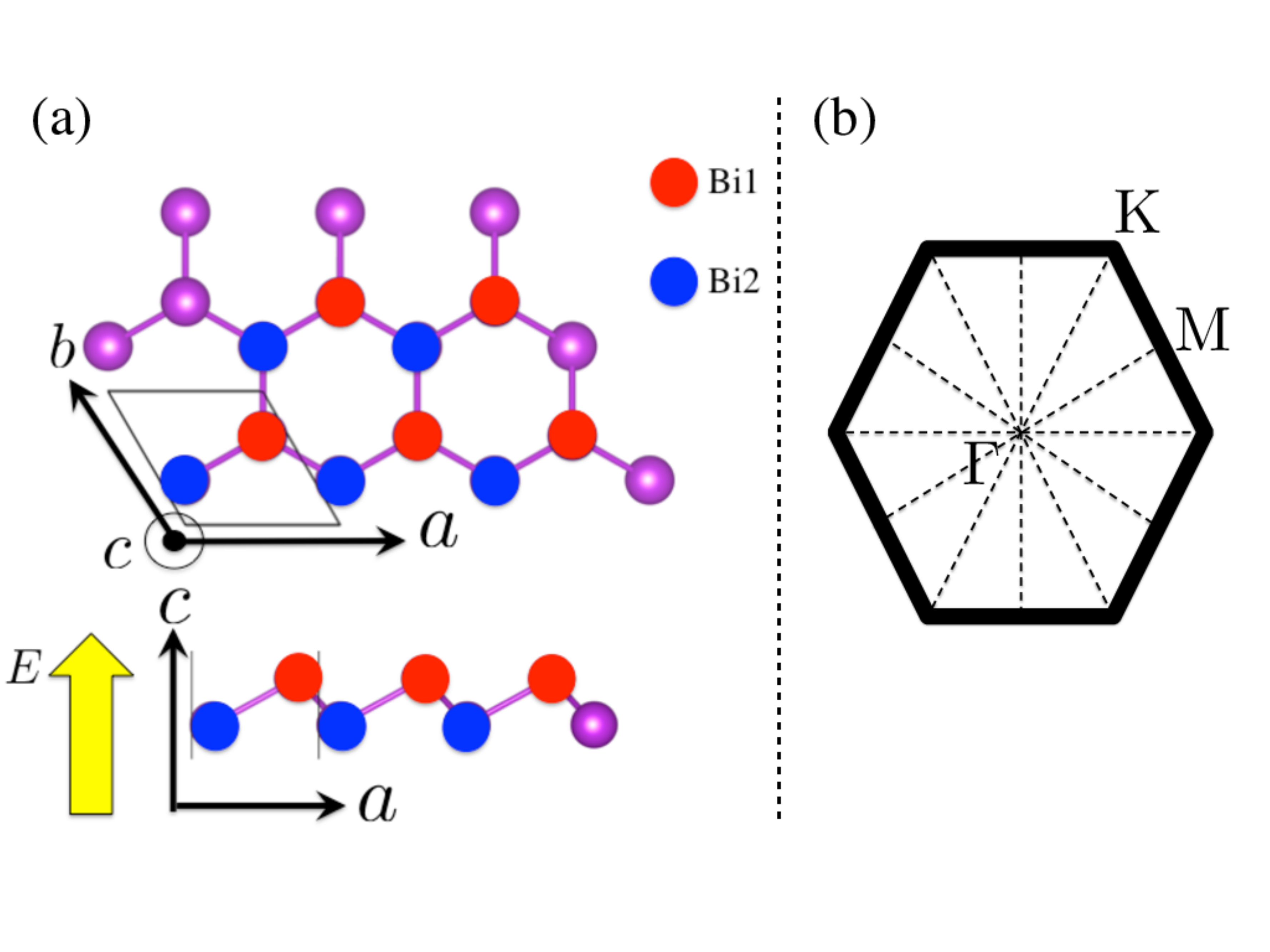}
\vspace{-1.5cm}
\caption{(a) Top view (upper part) and side view (lower part) of the atomic structure. (b) Brillouin zone of one-bilayer Bi(111).}
\label{Bifig}
\end{figure}

We show the structure of the one-bilayer Bi(111) film and the direction of the applied electric field in Fig. \ref{Bifig}(a). The unit cell is a honeycomb lattice, similar to that of graphene, with a buckled structure. We set the length of the side in the cell as 4.54 {\AA} and the buckling height as 1.45 {\AA}, as reported experimentally\cite{Nagao_Nanofilm_2004}.
\par

We perform fully relativistic density functional calculations using the OpenMX\cite{OpenMX} code. We use the local spin density approximation (LSDA)\cite{Perdew_Self_1981} as the exchange–correlation functional. We use norm-conserving pseudopotentials\cite{Hamann_Norm_1979} and the linear combination of multiple pseudoatomic orbitals (LCPAO) for wave function expansion\cite{Ozaki_Variationally_2003,Ozaki_Numerical_2004}. The cutoff radius is specified as 8.0 Bohr radii and the pseudoatomic orbitals are specified by $s3p3d2$ with three $s$-orbitals, three $p$-orbitals, and two $d$-orbitals. We set the cutoff energy of 300 Ry, and the ${\bf k}$-space sampling points of $13\times13\times1$ for the reciprocal lattice vectors. We include the spin-orbit interactions by a $j$-dependent pseudopotential composed relativistically (fully relativistic pseudopotential)\cite{Theurich_Self_2001}. We neglect changes in the lattice parameters and atomic positions induced by the electric field. The electric field is introduced as a sawtooth potential, and the system is calculated as a one-bilayer slab model. These computational conditions are the same as those in a previous study\cite{Kotaka_rashba}.\par

We implement methods for computing the $Z_{2}$ invariants by WFC and LCN in OpenMX\cite{OpenMX} code. $Z_{2} =1\ ({\rm mod}\ 2)$ corresponds to the topological insulator, and $Z_{2} = 0\ ({\rm mod}\ 2)$ corresponds to the trivial insulator. We compute the $Z_{2}$ invariant by the two methods applicable to systems without inversion symmetry.\par

\begin{figure}[htbp]
\includegraphics[width=\columnwidth]{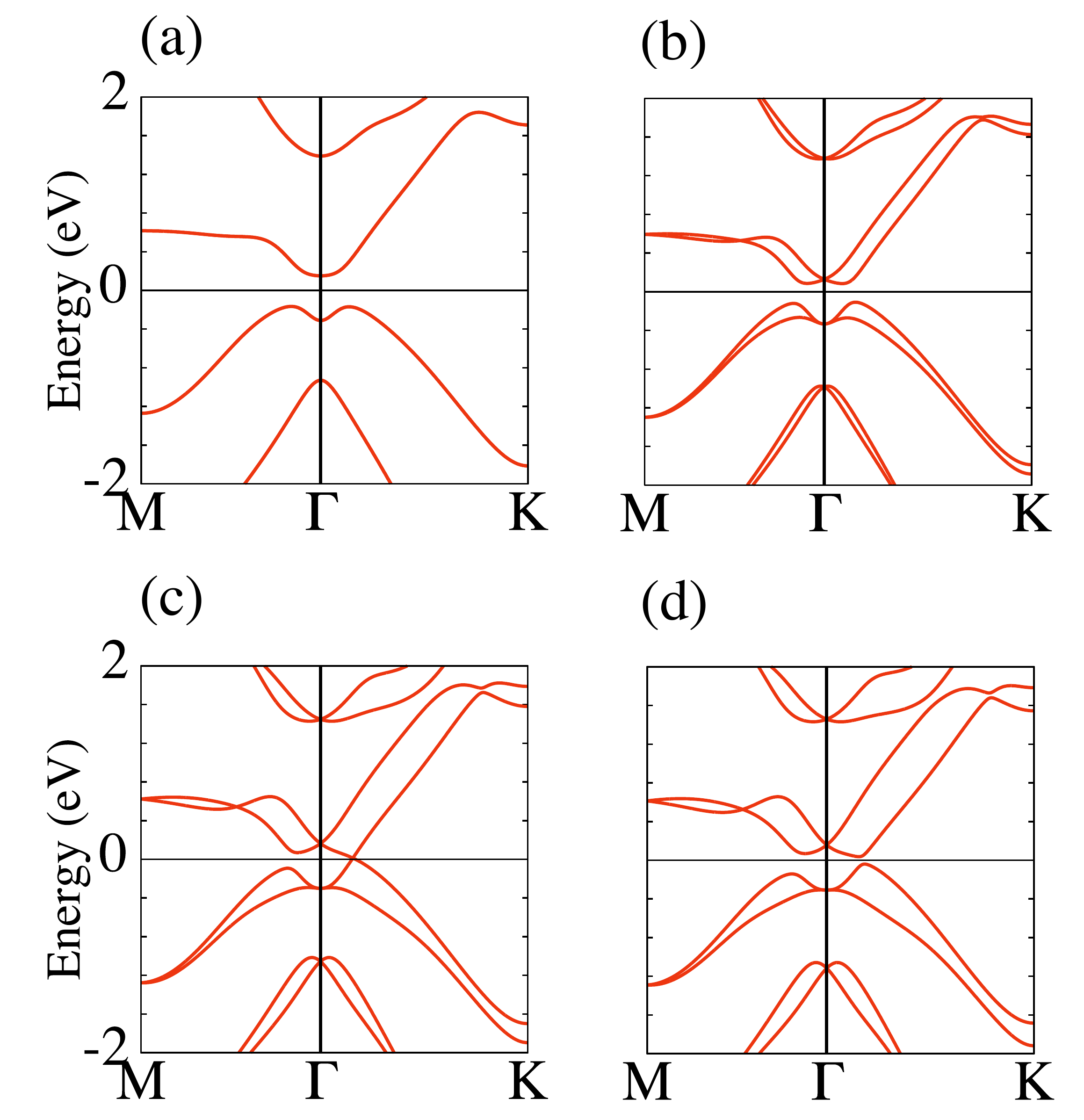}
\caption{Band structure of one-bilayer Bi(111) at
(a) $E=0\ {\rm V/\AA}$, topological insulator phase
(b) $E=1.0\ {\rm V/\AA}$, topological insulator phase (c) $E=2.1\ {\rm V/\AA}$, Dirac metal phase
(d) $E=2.5\ {\rm V/\AA}$, trivial insulator phase
}
\label{Bandfig}
\end{figure}

First, we investigate the electric field dependence of the band structure of the system. One-bilayer Bi(111) has time reversal symmetry 
$\epsilon({\bf k},\uparrow)=\epsilon(-{\bf k},\downarrow)$,
and in the absence of electric fields it has the space inversion symmetry 
$\epsilon({\bf k},\uparrow)=\epsilon(-{\bf k},\uparrow)$.
Thus, the system becomes doubly degenerate 
$\epsilon({\bf k},\uparrow)=\epsilon({\bf k},\downarrow)$
at general ${\bf k}$ points. 
The application of an electric field causes spin splitting and resolves the degenerate 
$\epsilon({\bf k},\uparrow)=\epsilon({\bf k},\downarrow)$
for general ${\bf k}$ points, as well as the degeneracy of eigenvalues 
$\epsilon({\bf k},\uparrow)\neq\epsilon(-{\bf k},\uparrow)$
except for the time-reversal invariant ${\bf k}$ point $\Gamma:\frac{2\pi}{a}(0, 0, 0)$ and ${\rm M}:\frac{2\pi}{a} (0.5, 0, 0)$. 
Figures \ref{Bandfig}(a)-\ref{Bandfig}(d) show the electric field dependence of the band dispersion for one-bilayer Bi(111) under fields of $E =0, 1.0, 2.1$, and 2.5 V/{\AA}, respectively. We confirm electric field-induced spin splitting as shown in Figs. \ref{Bandfig}(b) and \ref{Bandfig}(c). The bandgap is decreased from 0.32 eV to 0 eV when the applied electric field reaches 2.1 V/{\AA}. Then, the bandgap is closed at ${\bf k} = \frac{2\pi}{a} (0.053, 0.053, 0)$ as shown in Fig. \ref{Bandfig}(c). Further increasing the applied electric field opens the bandgap, which reaches 0.34 eV at 4.0 V/{\AA}. 
\par

Next, we investigate the $Z_{2}$ invariant of one-bilayer Bi(111) in the absence of an electric field. In a space-inversion symmetric system, the existence of the topological insulator phase can be distinguished by a product of the parity eigenvalues at the time-reversal invariant ${\bf k}$ points, 
$\frac{2\pi}{a}(0, 0, 0), \frac{2\pi}{a}(0.5, 0, 0),\frac{2\pi}{a}(0, 0.5, 0)$, and 
$\frac{2\pi}{a}(0.5, 0.5, 0)$. 
If the product is negative, $(-1)^{Z_{2}}={\displaystyle\prod_{i=1}^{4}}\delta_{i}$,
the system is topological insulator phase, and $Z_{2} = 1$\cite{Fu_Topological_2007}. 
Calculated $\delta_{i}$ are -1,+1,-1,-1 respectively, thus
we confirm that $Z_{2} = 1$ by the product of parity eigenvalues on one-bilayer Bi(111). This result agrees with the previous study\cite{Murakami_Quantum_2006}. \par

\begin{figure}[htbp]
\vspace{-0.5cm}
\includegraphics[width=\columnwidth]{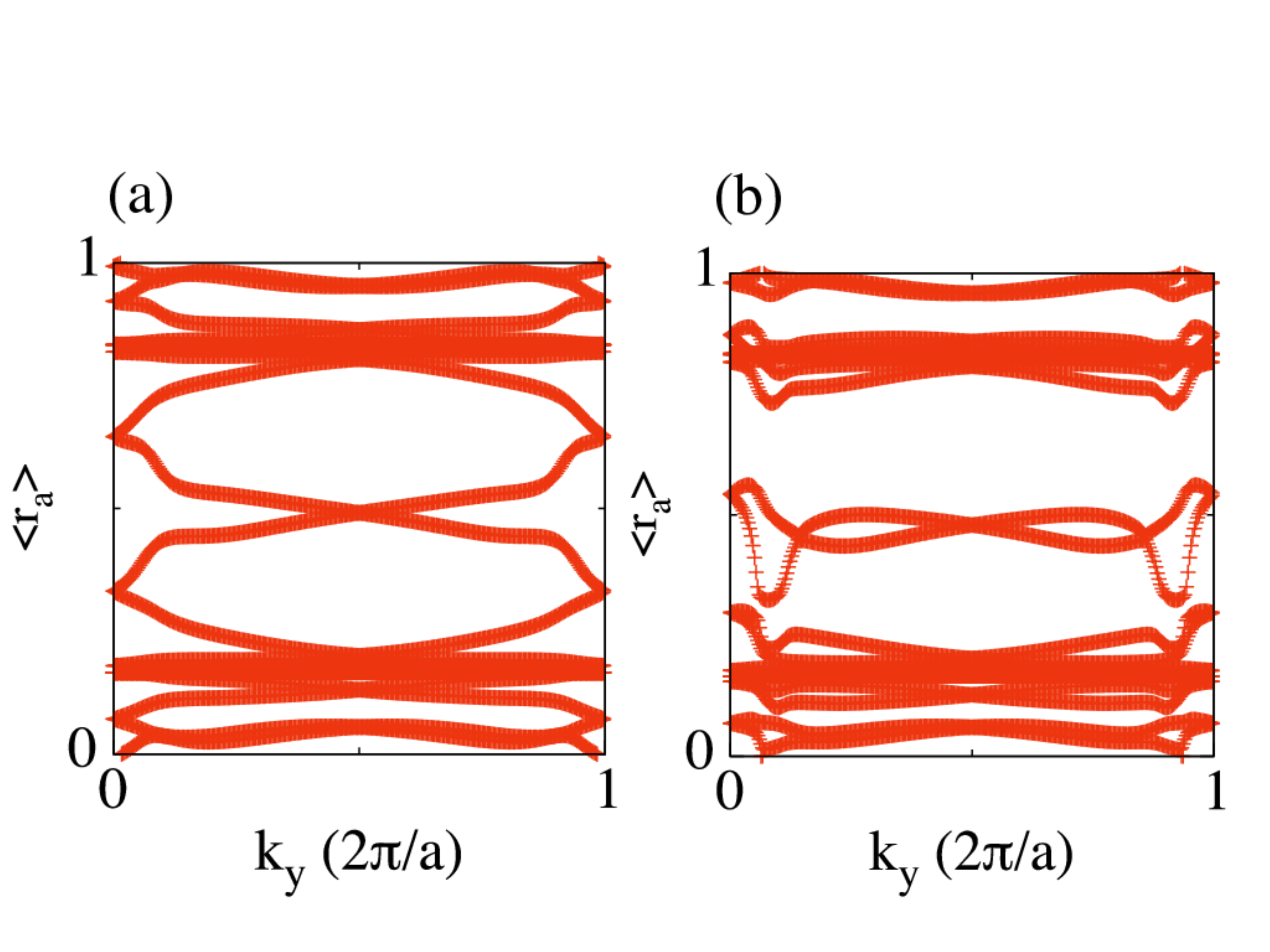}
\vspace{-1cm}
\caption{
The evolution of WFC lines, $k_{y}$ dependence of $\langle r_{a}\rangle$.
If these lines have no gaps, the system is topological insulator phase.
(a) $E=1.0\ {\rm V/\AA}$, topological insulator phase
(b) $E=2.5\ {\rm V/\AA}$, trivial insulator phase
}
\label{wfcfig}
\end{figure}

We compute the $Z_{2}$ invariant of one-bilayer Bi(111) under an applied electric field. Figure \ref{wfcfig} shows the $k_{y}$ dependence of occupied WFCs $\langle r_{a} \rangle$ along the $a$-axis. 
The WFCs are computed by the one-dimension hybrid Wannier functions\cite{Marzari_Maximally_1997}. 
The topological insulator phase can be distinguished from the trivial insulator phase by the evolution of the WFC lines which are the $k_{y}$ dependence of $\langle r_{a} \rangle$. If the lines show no gap, the system is topological insulator phase with $Z_{2} = 1$\cite{Soluyanov_Computing_2011,Yu_Equivalent_2011}. 
Figure \ref{wfcfig}(a) shows WFCs at $E=1.0$ V/{\AA}. The WFC lines cross at $k_{y} = 0, \pi$. Thus, the topological phase is non-trivial and $Z_{2} = 1$. Figure  \ref{wfcfig}(b) shows WFCs at $E =2.5$ V/{\AA}; the WFC lines have gaps, and therefore the system is a trivial insulator with $Z_{2} = 0$. According to these results, 2.1 V/{\AA} is the critical electric field strength to induce the topological phase transition from topological insulator phase to the trivial insulator phase. \par

We also compute the $Z_{2}$ invariants by the LCN to validate the results obtained by the WFC. Figure \ref{lcnfig} shows examples of LCN of the system with the applied electric fields of 1.0 V/{\AA} and 2.5 V/{\AA}. The unfilled circles correspond to LCN of +1, filled circles to LCN of -1, and blanks to LCN of 0. The topological insulator phase can be distinguished from the trivial insulator phase by a summation of the LCN in the half Brillouin zone (corresponding to white space in Fig. {\ref{lcnfig}). If the summation of LCN is an odd number, the system is topological insulator phase and $Z_{2} = 1$\cite{Fukui_Quantum_2007,Feng_First_2012}. These results are consistent with the results of WFC.\par
\begin{figure}[htbp]
\vspace{0.5cm}
\includegraphics[width=\columnwidth]{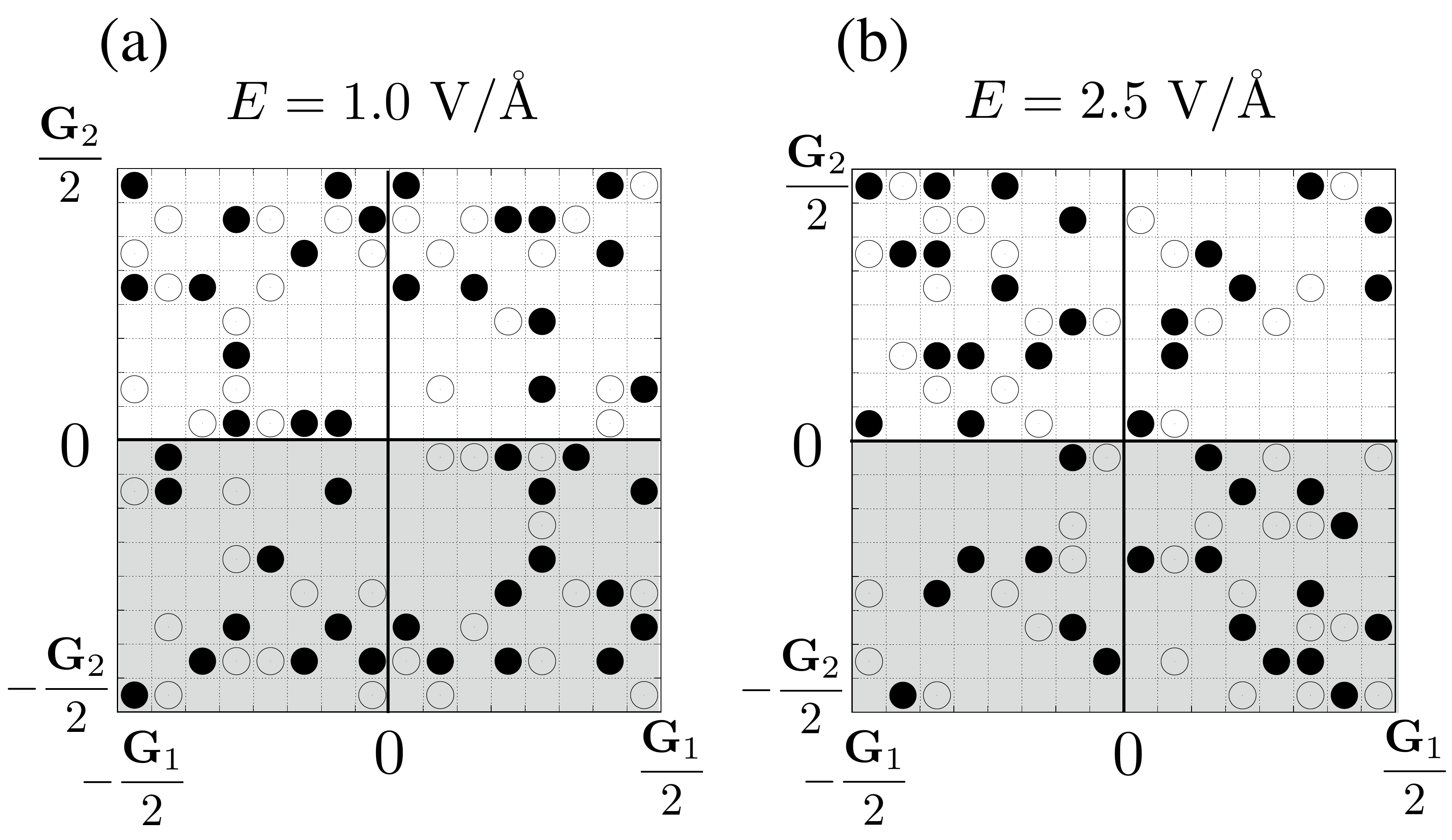}
\caption{The calculated  lattice Chern numbers (LCNs) in discretized Brillouin zone for  (a) $E = 1.0$ V/{\AA} and (b) $E = 2.5$ V/{\AA}.
Unfilled circles and filled circles indicate LCNs of +1 and -1, respectively.
$Z_{2}$ topological invariant is computed as the total LCNs modulo two in half Brillouin zone, gray or white area.}
\label{lcnfig}
\end{figure}


We confirm the appearance of a Dirac metal state at the topological phase transition. Figure \ref{Diracmetal}(a) shows the band structure at $E = 2.1$ V/{\AA}. We confirm the Dirac cone at ${\bf k} =\frac{2\pi}{a} (0.053, 0.053, 0)$. 
In the Dirac cone, the Berry flux 
$\Phi=\int {\bf A}\cdot d{\bf l},{\bf A}=\bra{u_{\bf k}}
\frac{\partial}{\partial {\bf k}} \ket{u_{\bf k}}$ 
becomes 
$\Phi=\pi$.
We investigate the electric field dependence of the Berry flux\cite{doi:10.1143/JPSJ.74.1674} around ${\bf k}=\frac{2\pi}{a}(0.053,0.053,0)$ in Fig. \ref{Diracmetal}(b), which indicates the Dirac metal state.\par
\begin{figure}[htbp]
\includegraphics[width=\columnwidth]{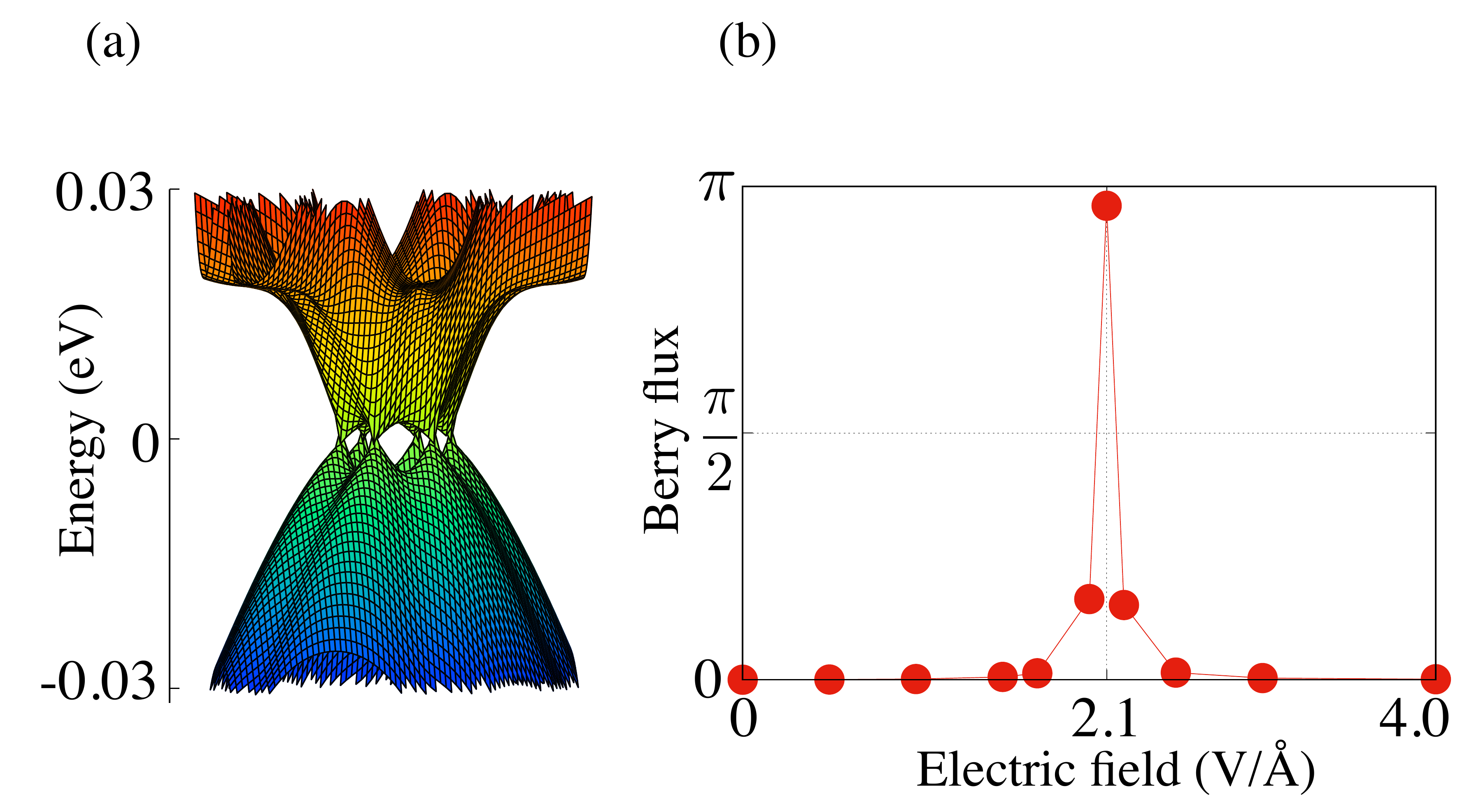}
\caption{(a) Three-dimensional plot of Dirac cone at $E=2.1$ V/{\AA}. (b) The electric field dependence of the Berry flux around ${\bf k}=\frac{2\pi}{a}$(0.053,0.053,0).}
\label{Diracmetal}
\end{figure}


We show the topological phase diagram, the electric field dependence of $Z_{2}$ invariant, and the bandgap in Fig. \ref{phasediagram}. 
One-bilayer Bi(111) transitions from the topological insulator phase to the trivial insulator phase via a Dirac metal state. These results are consistent with the topological insulator phase of the system remaining for $E < 0.8$ V/{\AA}\cite{Chen_Robustness_2013} and with the monotonic decrease of the bandgap for $E < 1.5$ V/{\AA}\cite{Kotaka_rashba}. We predicted a strong electric field-induced topological phase transition in one-bilayer Bi(111).\par

\begin{figure}[htbp]
\includegraphics[width=\columnwidth]{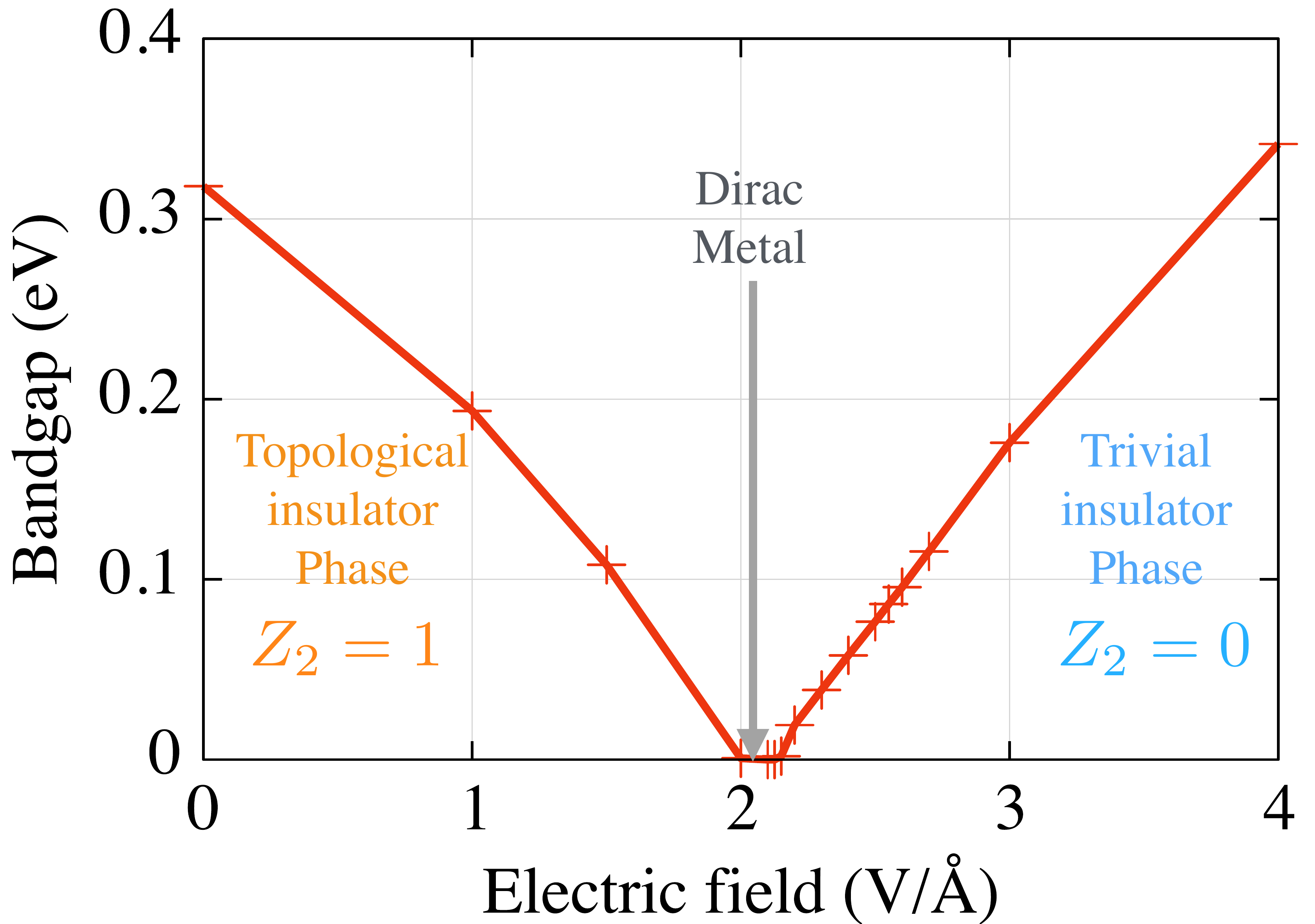}
\caption{Electric field dependence of the bandgap and  topological phase diagram.
}
\label{phasediagram}
\end{figure}


In order to understand the origin of the topological phase transition, we investigate the partial density of states (PDOS) and the wave function around the Dirac point. The topological phase transition is induced by the band inversion of two different characters. We plot the PDOS near the Fermi level in Fig. \ref{PDOS}. Comparing these PDOS, the inversion of PDOS is apparent for Atoms Bi1 and Bi2 near the Fermi level. Furthermore, we plot the Bloch wave function of the valence band and conduction band at ${\bf k} = \frac{2\pi}{a} ( 0.05, 0.05, 0)$ in Fig. \ref{wavefunc}. Comparing the wave functions, we confirm the inversion of the wave function; one spreads along the (100) direction and the other along the (110) direction. This inversion of wave functions arises from charge transfer between Atoms Bi1 and Bi2 by the applied electric field. Thus, we demonstrate that the topological phase transition in one-bilayer Bi(111) is induced by the band inversion of the valence and conduction bands.\par

\begin{figure}[htbp]
\includegraphics[width=\columnwidth]{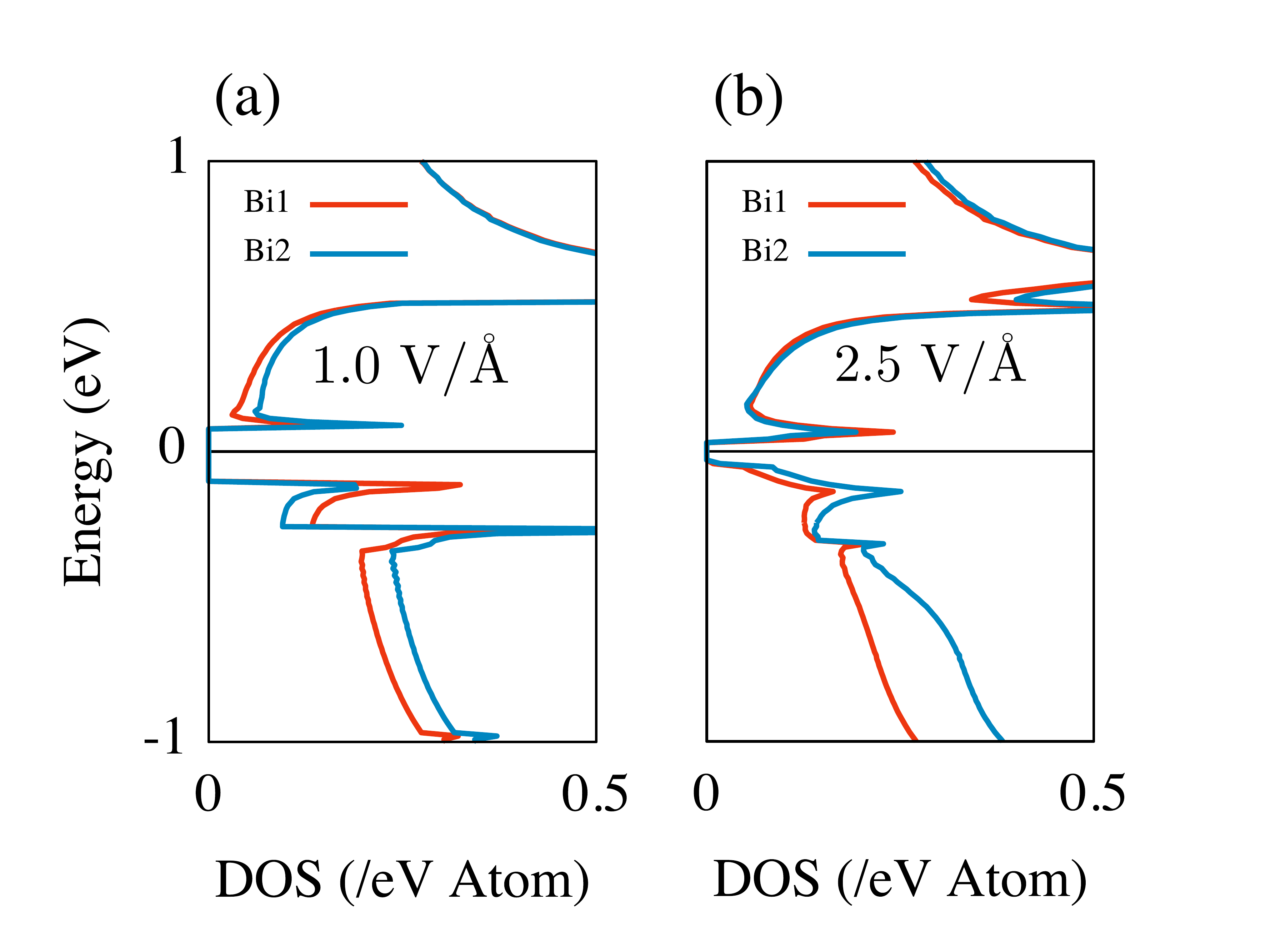}
\caption{The partial density states of Bi1 and Bi2 at (a) $E=1.0$ V/{\AA} and (b) $E= 2.5$ V/{\AA}. }
\label{PDOS}
\end{figure}

\begin{figure}[htbp]
\includegraphics[width=\columnwidth]{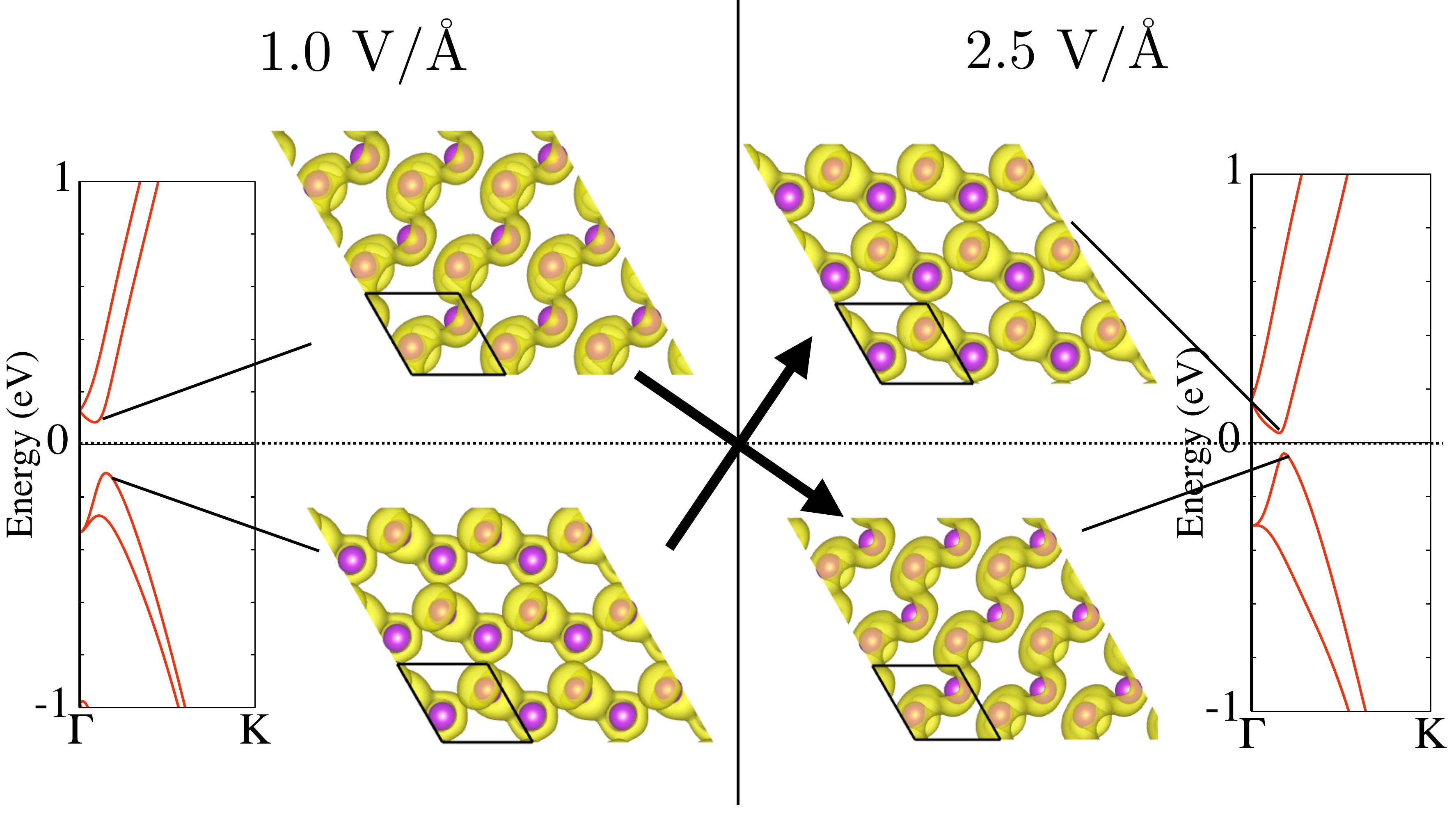}
\caption{
The electric-field induced band inversion of  valence band and conduction band at ${\bf k}=\frac{2\pi}{a}(0.05,0.05,0)$.
The band structure and  isosurface of electron wave functions $|\psi_{\bf k}|^2$ are plotted for topological insulator phase ($E=1.0$ V/{\AA}) and trivial insulator phase ($E=2.5$ V/{\AA}).}
\label{wavefunc}
\end{figure}


In summary, we computed the electric field dependence of the bandgap and $Z_{2}$ invariant of one-bilayer Bi(111). We predicted that  topological phase transitions from a topological insulator phase to a trivial insulator phase for $E > 2.1$ V/{\AA} one-bilayer Bi(111). We plotted the PDOS and wave function around the Dirac point, and we could see the band character inversion. Thus, we confirmed that the topological phase transition was induced by band inversion in in one-bilayer Bi(111).
In considering topological-to-trivial switching devices, $E = 2.1$ V/{\AA} is a very strong electric field. However, this critical electric field is proportional to spin-orbit interactions; it could be reduced by alloying atoms with smaller spin-orbit interactions than Bi, such as Sb or As.
\acknowledgment
This work was supported by Grant-in-Aid for Scientific Research on Innovative Area, ”Nano Spin Conversion Science” (Grant Nos. 15H01015 and 17H05180).
The work was partially supported by Grants-in-Aid on Scientific Research under Grant No. 16K04875 from Japan Society for the Promo-
tion of Science.
The computations in this research were performed using the supercomputers at
 RIIT, Kyushu University, and the ISSP, University of Tokyo.


\bibliographystyle{jjap}
\bibliography{reference}

\providecommand{\noopsort}[1]{}\providecommand{\singleletter}[1]{#1}%
\begin{thebibliography}{10}
\expandafter\ifx\csname url\endcsname\relax
  \def\url#1{\texttt{#1}}\fi
\expandafter\ifx\csname urlprefix\endcsname\relax\def\urlprefix{URL }\fi

\bibitem{Hasan_Colloquium_2010}
M.~Hasan and C.~Kane, Rev. Mod. Phys. {\bf 82}, 3045 (2010) .

\bibitem{Liu_Switching_2015}
Q.~Liu, X.~Zhang, L.~Abdalla, A.~Fazzio and A.~Zunger, Nano Lett. {\bf 15},
  1222 (2015) .

\bibitem{Nagao_Nanofilm_2004}
T.~Nagao, J.~Sadowski, M.~Saito, S.~Yaginuma, Y.~Fujikawa, T.~Kogure, T.~Ohno,
  Y.~Hasegawa, S.~Hasegawa and T.~Sakurai, Phys. Rev. Lett. {\bf 93}, 10 (2004)
  .

\bibitem{Murakami_Quantum_2006}
S.~Murakami, Phys. Rev. Lett. {\bf 97}, 236805 (2006) .

\bibitem{Wada_Localized_2011}
M.~Wada, S.~Murakami, F.~Freimuth and G.~Bihlmayer, Phys. Rev. B {\bf 83},
  121310 (2011) .

\bibitem{Hirahara_Interfacing_2011}
T.~Hirahara, G.~Bihlmayer, Y.~Sakamoto, M.~Yamada, H.~Miyazaki, S.-i. Kimura,
  S.~Bl{\"u}gel and S.~Hasegawa, Phys. Rev. Lett. {\bf 107}, 166801 (2011) .

\bibitem{Yao_Topologically_2016}
M.~Yao, F.~Zhu, C.~Han, D.~Guan, C.~Liu, D.~Qian and J.-f. Jia, Sci. Reports
  {\bf 6}, 21326 (2016) .

\bibitem{Chen_Robustness_2013}
L.~Chen, Z.~Wang and F.~Liu, Phys. Rev. B {\bf 87}, 23 (2013) .

\bibitem{Kotaka_rashba}
H.~Kotaka, F.~Ishii and M.~Saito, Jpn. J. App. Phys. {\bf 52}, 035204 (2013) .

\bibitem{Soluyanov_Computing_2011}
A.~A. Soluyanov and D.~Vanderbilt, Phys. Rev. B {\bf 83}, 235401 (2011) .

\bibitem{Yu_Equivalent_2011}
R.~Yu, X.~Qi, A.~Bernevig, Z.~Fang and X.~Dai, Phys. Rev. B {\bf 84}, 075119
  (2011) .

\bibitem{Fukui_Quantum_2007}
T.~Fukui and Y.~Hatsugai, J. Phys. Soc. Jpn. {\bf 76}, 053702 (2007) .

\bibitem{Feng_First_2012}
W.~Feng, J.~Wen, J.~Zhou, D.~Xiao and Y.~Yao, Comput. Phys. Commun. {\bf 183},
  1849 (2012) .

\bibitem{Fu_Topological_2007}
L.~Fu and C.~Kane, Phys. Rev. B {\bf 76}, 045302 (2007) .

\bibitem{OpenMX}
T.~Ozaki~et al., \url{http://www.openmx-square.org/}.

\bibitem{Perdew_Self_1981}
J.~Perdew and A.~Zunger, Phys. Rev. B {\bf 23}, 5048 (1981) .

\bibitem{Hamann_Norm_1979}
D.~Hamann, M.~Schl{\"u}ter and C.~Chiang, Phys. Rev. Lett. {\bf 43}, 1494
  (1979) .

\bibitem{Ozaki_Variationally_2003}
T.~Ozaki, Phys. Rev. B {\bf 67}, 15 (2003) .

\bibitem{Ozaki_Numerical_2004}
T.~Ozaki and H.~Kino, Phys. Rev. B {\bf 69}, 19 (2004) .

\bibitem{Theurich_Self_2001}
G.~Theurich and N.~A. Hill, Phys. Rev. B {\bf 64}, 7 (2001) .

\bibitem{Marzari_Maximally_1997}
N.~Marzari and D.~Vanderbilt, Phys. Rev. B {\bf 56}, 12847 (1997) .

\bibitem{doi:10.1143/JPSJ.74.1674}
T.~Fukui, Y.~Hatsugai and H.~Suzuki, J. Phys. Soc. Jpn. {\bf 74}, 1674 (2005) .

\end{thebibliography}

\end{document}